\documentclass[preprint,12pt]{elsarticle}

\usepackage{amssymb}

\usepackage[dvipsnames]{xcolor}
\usepackage[utf8]{inputenc}
\usepackage[T1]{fontenc}
\usepackage{array,multirow}
\usepackage{comment}

\usepackage{natbib}
\biboptions{sort&compress}

\usepackage{lineno}

\usepackage[
  letterpaper,
  left=2.5cm,
  right=2.5cm
]{geometry}

\usepackage{amsmath}

\journal{Physical Review Fluids}

\newcommand\Deb{\mbox{\textit{De}}} 
\newcommand\Wei{\mbox{\textit{Wi}}} 
\newcommand\Oh{\mbox{\textit{Oh}}}  
\newcommand\Bo{\mbox{\textit{Bo}}}  

\begin{document}

\begin{frontmatter}



\title{Beware of CaBER: Filament thinning rheometry does not always give `the' relaxation time of polymer solutions}



\cortext[cor]{Corresponding author}

\author[aff1]{A. Gaillard}
\ead{antoine0gaillard@gmail.com}
\author[aff2]{M. A. Herrada}
\author[aff1]{A. Deblais}
\author[aff3]{J. Eggers}
\author[aff1]{D. Bonn}

\affiliation[aff1]{organization={Van der Waals-Zeeman Institute, University of Amsterdam}, 
            addressline={Science Park 904}, 
            city={Amsterdam},
            country={Netherlands}}
            
\affiliation[aff2]{organization={Depto. de Mecánica de Fluidos e Ingeniería Aeroespacial, Universidad de Sevilla}, 
            city={Sevilla},
            postcode={E-41092}, 
            country={Spain}}

\affiliation[aff3]{organization={School of Mathematics, University of Bristol}, 
            addressline={University Walk}, 
            city={Bristol},
            postcode={BS8 1 TW}, 
            country={United Kingdom}}

\begin{abstract}
The viscoelastic relaxation time of a polymer solution is often measured using Capillary Breakup Extensional Rheometry (CaBER) where a droplet is placed between two plates which are pulled apart to form a thinning filament. For a slow plate retraction protocol, required to avoid inertio-capillary oscillations for low-viscosity liquids, we show experimentally that the CaBER relaxation time $\tau_e$ inferred from the exponential thinning regime is in fact an apparent relaxation time that may increase significantly when increasing the plate diameter and the droplet volume. Similarly, we observe that $\tau_e$ increases with the plate diameter for the classical step-strain plate separation protocol of a commercial (Haake) CaBER device and increases with the nozzle diameter for a Dripping-onto-Substrate (DoS) method. This dependence on the flow history before the formation of the viscoelastic filament is in contradiction with polymer models such as Oldroyd-B that predict a filament thinning rate $1/3\tau$ ($\tau$ being the model's relaxation time) which is a material property independent of geometrical factors. We show that this is not due to artefacts such as solvent evaporation or polymer degradation and that it can only be rationalised by finite extensibility effects (FENE-P model) for a dilute polymer solution in a viscous solvent, but not for semi-dilute solutions in a low-viscosity solvent.
\end{abstract}




\end{frontmatter}

\section{Introduction}

When polymers are added to a low-viscosity solvent such as water, the extensional rheology of the resulting solution is usually measured by indirect techniques where the (extensional) strain and strain rate are not controlled, unlike for high-viscosity polymer solutions or melts for which reliable extensional rheometers are available, e.g. Meissner's RME (Rheometric Melt Elongation rheometer) and FiSER (Filament Stretching Extensional Rheometer). Most indirect techniques for low-viscosity polymer solutions aim at forming a liquid filament undergoing capillary-driven thinning. Historically, this was first achieved by placing a drop of liquid between two horizontal plates which are then separated beyond the stability limit of a stable liquid bridge \cite{bazilevsky1997failure,anna2001elasto,stelter2000validation}, a technique now known as CaBER (Capillary Breakup Extensional Rheometry). Alternative techniques, also based on the Rayleigh-Plateau instability, were proposed to avoid inertio-capillary oscillations of the end-drops in the original CaBER step-strain (rapid) plate separation protocol which prohibits measurement of very short relaxation times \cite{rodd2005capillary}. This is achieved by separating the plates at a constant low velocity (Slow Retraction Method or SRM) \cite{campo2010slow}, by dripping a droplet in air from a nozzle at a low flow rate \cite{tirtaatmadja2006drop} or, in a more recent technique, by slowly bringing a solid substrate in contact with a drop hanging steadily from a nozzle (Dripping-onto-Substrate or DoS) \cite{dinic2017pinch}.

In all these techniques, after an initial inertial and/or viscous regime, an elastic regime emerges where the elastic stresses arising from the stretching of polymer chains dominate and give rise to a cylindrical filament that thins exponentially in time for a wide range of dilute and semi-dilute polymer solutions. This is consistent with the Oldroyd-B model that predicts \cite{bazilevsky1990liquid,entov1997effect}: 
\begin{equation}
    h = h_1 \exp{\left( -\frac{t-t_1}{3\tau} \right)} \,\mathrm{,}
\label{eq:exponential}
\end{equation}
\noindent where $h$ is the minimum filament radius, $\tau$ the viscoelastic relaxation time of the polymer solution, $t_1$ the time marking the onset of the elastic regime, and $h_1=h(t_1)$ the filament radius at that time. For a step-strain CaBER protocol, in which polymer molecules do not relax during the fast plate separation, the model predicts $h_1 = (G h_i^4 / 2 \gamma)^{1/3}$ where $G = \eta_p/\tau$ is the elastic modulus, $\eta_p$ the polymer contribution to the shear-viscosity, $\gamma$ the surface tension and $h_i$ the radius of the initial liquid column \cite{clasen2006beads}. It is generally accepted that (i) for a polymer solution with a spectrum of relaxation times, the longest one dominates \cite{anna2001elasto} and that (ii) as polymer chains unravel during the exponential regime, they ultimately approach their finite extensibility limit, causing the filament to break after a terminal regime which can be described by, e.g., FENE models (P or CR) \cite{entov1997effect,anna2001elasto,clasen2006dilute}.

The general consensus is that geometrical parameters such as the size of the system can only influence $h_1$ (via $h_i$) but not the thinning rate $\vert \dot{h}/h \vert = 1/3 \tau$ of the filament (where the dot means $\mathrm{d}/ \mathrm{d}t$) since $\tau$ is a material property. In particular, Bazilevsky et al. \cite{bazilevsky1997failure} and Miller et al. \cite{miller2009effect} checked that the filament thinning rate was independent of the sample volume and of the plate separation speed and, in a step-strain plate separation protocol, on the final plate separation distance. This suggests that it is independent of the history of the polymer deformation prior to the elastic regime. However, Rajesh et al. \cite{rajesh2022transition} recently tested polymer solutions of different solvent viscosities with a dripping method and reported a larger thinning rate for a smaller nozzle radius.

In this paper, we show that the apparent relaxation time, inferred from the exponential thinning regime, depends on the size of the system for other filament thinning techniques such as CaBER (with both slow and fast plate separation protocols) and Dripping-onto-Substrate (DoS).

\section{Materials and methods}
\label{sec:Materials and methods}

\subsection{Polymer solutions}
\label{subsec:Polymer solutions}

\noindent We use three different liquids: two solutions of poly(ethylene oxide) (PEO) of molecular weight $M_w = 4 \times 10^{6}$~g/mol, one in water with concentration $500$~(w)ppm, referred to as PEO$_{\mathrm{aq}}$, and one in a more viscous solvent with concentration $25$~(w)ppm, referred to as PEO$_{\mathrm{visc}}$, and a $1000$~ppm solution of poly(acrylamide/sodium acrylate) (HPAM) [70:30] of molecular weight $M_w = 18 \times 10^{6}$~g/mol in water with $1$~wt\% NaCl to screen electrostatic interactions and make the chain flexible instead of semi-rigid. Both polymers were provided by Polysciences (ref. 04030-500 for PEO and 18522-100 for HPAM). For the PEO$_{\mathrm{visc}}$ solution, the solvent is an aqueous Newtonian $30$~wt\% $20,000$~g/mol PEG solution. The different concentrations were chosen to ensure that all three liquids have comparable filament thinning rates. After slowly injecting the polymer powder to a vortex generated by a magnetic stirrer, solutions were homogenised using a mechanical stirrer at low rotation speed for about $16$ hours. For the PEO$_{\mathrm{visc}}$ solution, PEG was added after mixing PEO with water.

\begin{table}
\setlength{\tabcolsep}{2.5pt}
  \begin{center}
  \begin{footnotesize} 
\def~{\hphantom{0}}
  \begin{tabular}{ccccccccccc}
  
    Name~~                & $\rho$     & $\gamma$ & $\eta_s$ & $c$   & $c/c^*$ & $\eta_0$ & $\eta_p$ & $n$  & $1/\dot{\gamma}_c$ & $\tau_{m}$ \\
                          & (kg/m$^3$) & (mN/m)   & (mPa~s)  & (ppm) &         & (mPa~s)  & (mPa~s)  &      & (ms)               & (ms) \\ [8pt]      
    PEO$_{\mathrm{aq}}$~  & 998        & 62.5     & 0.92     & 500   & 1.86    & 3.0      & 2.08     & 0.93 & 120 &   240 \\
    PEO$_{\mathrm{visc}}$ & 1048       & 56.0     & 245      & 25    & 0.018   & 248      & 3.3      & 1    & --    & 110 \\
    HPAM~                 & 998        & 72.0     & 0.92     & 1000  & --      & 15       & 14       & 0.78 & 410  &  100 \\
  \end{tabular}
  \caption{Properties of the three polymer solutions. $\rho$ is the density, $\gamma$ the surface tension, $\eta_s$ the solvent viscosity, $c$ the polymer concentration, $c^*$ the critical overlap concentration, $\eta_0$, $n$ and $\dot{\gamma}_c$ the Carreau fitting parameters of the shear viscosity, $\eta_p = \eta_0-\eta_s$, and $\tau_{m}$ the maximum CaBER relaxation time measured for the largest plates.} 
  \label{tab:rheology_shortpaper}
  \end{footnotesize} 
  \end{center}
\end{table}

The shear viscosity $\eta$ of these solutions was measured at the temperature of filament thinning experiments with a MRC-302 rheometer from Anton Paar equipped with a cone plate geometry (diameter $50$~mm, angle $1^{\circ}$ and truncation gap $53$~$\mu$m). The PEO$_{\mathrm{visc}}$ solution is a Boger fluid with a constant shear viscosity while the two others are shear-thinning and are well described by the Carreau law $\eta(\dot{\gamma}) = \eta_0 ( 1 + ( \dot{\gamma}/\dot{\gamma}_c)^{2})^{(n-1)/2}$ where $\eta_0$ is the zero-shear viscosity, $n$ is the shear-thinning exponent and $\dot{\gamma}_c$ is the shear rate marking the onset of shear thinning. These values, along with the solvent viscosity $\eta_s$, the density $\rho$ and the surface tension $\gamma$ measured with a pendent drop method, are reported in table \ref{tab:rheology_shortpaper}. For the PEO$_{\mathrm{aq}}$ (500~ppm) solution, viscosity measurements for other PEO concentrations gave an intrinsic viscosity $[\eta] = 2.87$~m$^3$/kg and hence a critical overlap concentration $c^* = 0.77/[\eta] = 0.268$~kg/m$^3$ ($268$~ppm). Assuming that the PEO$_{\mathrm{visc}}$ solution (25~ppm) is dilute, $\eta_p$ should increase linearly with the concentration as $\eta_p = [\eta] \eta_s c$, from which $[\eta]$ and $c^*$ are estimated from this single PEO concentration. Values of $c/c^*$ are presented in table \ref{tab:rheology_shortpaper}.

\subsection{Slow stepwise plate separation CaBER protocol}
\label{subsec:Slow stepwise plate separation CaBER protocol}

\begin{figure}
  \centerline{\includegraphics[scale=0.57]{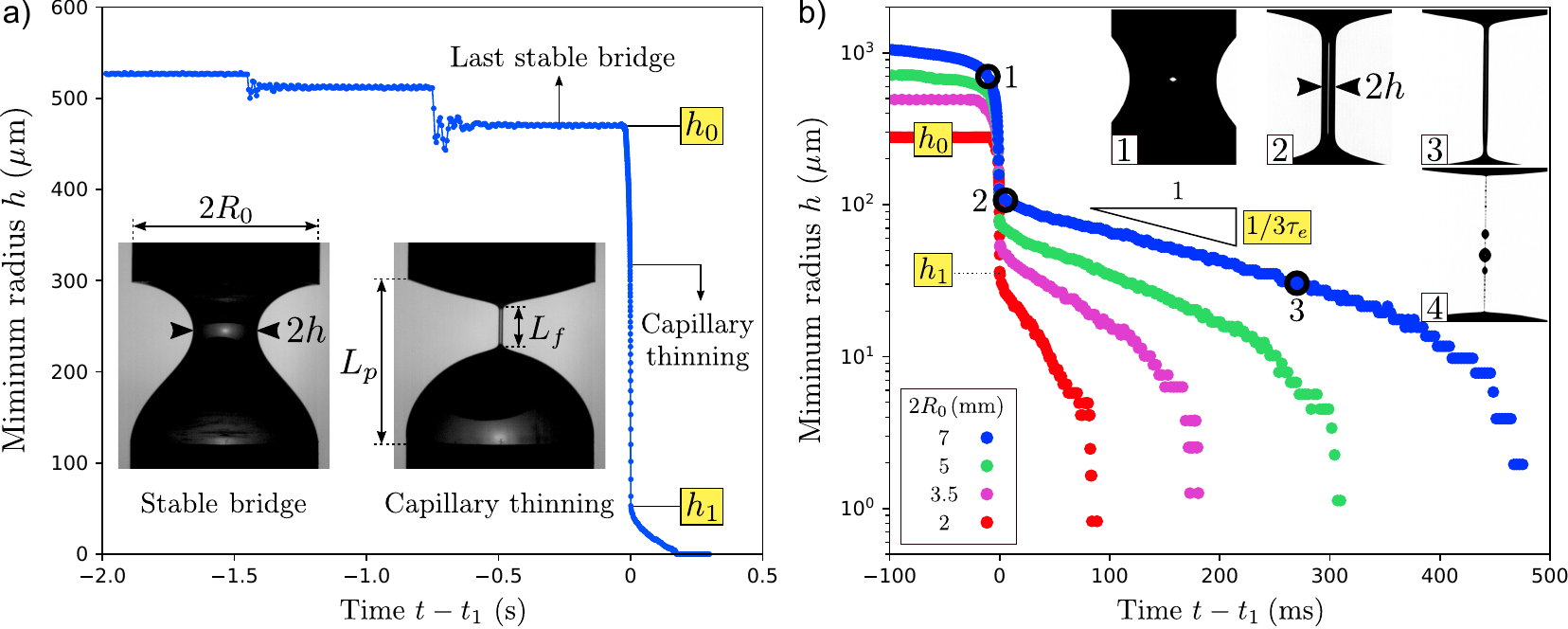}}
  \caption{(a) Time evolution of the minimum bridge / filament radius $h$ in the slow stepwise plate separation protocol for the PEO$_{\mathrm{aq}}$ solution for plate diameters $2R_0 = 3.5$~mm and a droplet volume $V^* = V/R_0^3 \approx 2.4$. Inertio-capillary oscillations are visible after each step. Inset images correspond to a stable liquid bridge (left) and to a thinning filament (right) of the PEO$_{\mathrm{aq}}$ solution with $2R_0 = 7$~mm and $V^* \approx 2.4$. (b) $h(t)$ in log-lin for the PEO$_{\mathrm{aq}}$ solution tested with plate diameters, $2R_0 = 2$, $3.5$, $5$ and $7$~mm, with $V^* \approx 2.4$. Inset images correspond to three times labelled 1 to 3 indicated on the $h(t)$ curve plus a fourth later time where $h$ is below our spatial resolution for $2R_0 = 7$~mm. The time reference $t_1$ marks the onset of the elastic regime. }
\label{fig:protocol}
\end{figure}

In our home-made CaBER setup, a droplet of volume $V$ is placed on a horizontal plate of radius $R_0$ and the motor-controlled top plate of same radius is first moved down until it is fully wetted by the liquid, i.e., until the liquid bridge between the plates has a quasi cylindrical shape. The top plate is then moved up slowly (at about $0.5$~mm/s) and stopped at a plate separation distance $L_p$ where the liquid bridge is still stable, like in the left inset image of figure \ref{fig:protocol}(a), but close to the bridge instability threshold. Then, instead of moving the top plate at a constant (lower) velocity, i.e. like in SRM \cite{campo2010slow}, we move it by 10~$\mu$m $L_p$-increment steps, waiting about one second between each step (longer than the solution's relaxation time), which is long enough to ensure that polymers are at equilibrium (no pre-stress) before each new step. At a certain step, the bridge becomes unstable and collapses under the action of surface tension, transiently leading to the formation of a nearly cylindrical filament which is the signature of viscoelastic pinch-off, as shown in the right inset image of figure \ref{fig:protocol}(a). We stop moving the top plate once capillary-driven thinning starts. The CaBER setup is placed in a plastic box where the relative humidity is kept above $80$\% using wet paper to minimise evaporation. The aluminium plates are plasma-treated before each new experiment to increase their hydrophilicity and minimise dewetting.

The process is recorded by a high-magnification objective mounted on a high-speed camera (Phantom TMX 7510) and images are analysed by a python code. A typical time-evolution of the minimum bridge / filament radius $h$ is shown in figure \ref{fig:protocol}(a). The purpose of this step-by-step plate separation protocol is to extract the value of the last stable bridge radius $h_0$ which, since steps are small, can be considered as the initial bridge radius at the onset of capillary thinning. Our image resolution is up to 1 pixel per micrometer for the smallest drops, corresponding to the smallest plates, and our time resolution is $15,000$ images per second to capture the fast bridge collapse from radius $h_0$ to the radius $h_1$ marking the onset of the elastic regime, see figure \ref{fig:protocol}(a).

The critical aspect ratio $\Lambda = L_p/(2 R_0)$ at which the liquid bridge becomes unstable depends on the liquid volume $V$ and on the Bond number $\Bo = \rho g R_0^2 / \gamma$, where $g$ is the gravitational acceleration \cite{slobozhanin1993stability}. In our experiments, we vary both the plate diameter $2R_0$, between $2$ and $25$~mm, and the non-dimensional droplet volume $V^* = V/R_0^3$ and we find that the last stable bridge radius $h_0$ increases with both $R_0$ and $V^*$.

\section{Results}
\label{sec:Results}

\noindent Figure \ref{fig:protocol}(b) compares the time-evolution of the minimum bridge / filament radius $h$ for the PEO$_{\mathrm{aq}}$ solution tested with plate diameters $2R_0$ between $2$ and $7$~mm with a fixed non-dimensional droplet volume $V^* = V/R_0^3 \approx 2.4$. Although all filaments thin exponentially in time at the beginning of the elastic regime, as suggested by the fairly straight curves for $t>t_1$ (before the terminal regime), they thin faster for smaller plates. This is in apparent contradiction with the Oldroyd-B model, which predicts a rate of exponential thinning $\vert \dot{h}/h \vert = 1/3 \tau$ (see equation \ref{eq:exponential}) which should be the same for all filaments, provided that the liquid does not change so that the (longest) relaxation time $\tau$ of the polymer solution is the same. 

To quantify these differences, we introduce an apparent (or effective) relaxation time $\tau_e$ such that $\vert \dot{h}/h \vert = 1/3 \tau_e$ in the exponential part of the elastic regime. As figure \ref{fig:protocol}(b) suggests, $\tau_e$ increases as the plate diameter increases. We show similar results for a Dripping-onto-Substrate (DoS) method in \ref{appA}, where $\tau_e$ is found to increase with the nozzle diameter. In \ref{appB}, we also show similar results for the classical step-strain plate separation protocol of a commercial (Haake) CaBER device where $\tau_e$ is found to increase with the plate diameter. This suggests a universal physical mechanism for the dependence of the filament thinning rate on the size of the system, independent of the exact method used.

\begin{figure}
  \centerline{\includegraphics[scale=0.57]{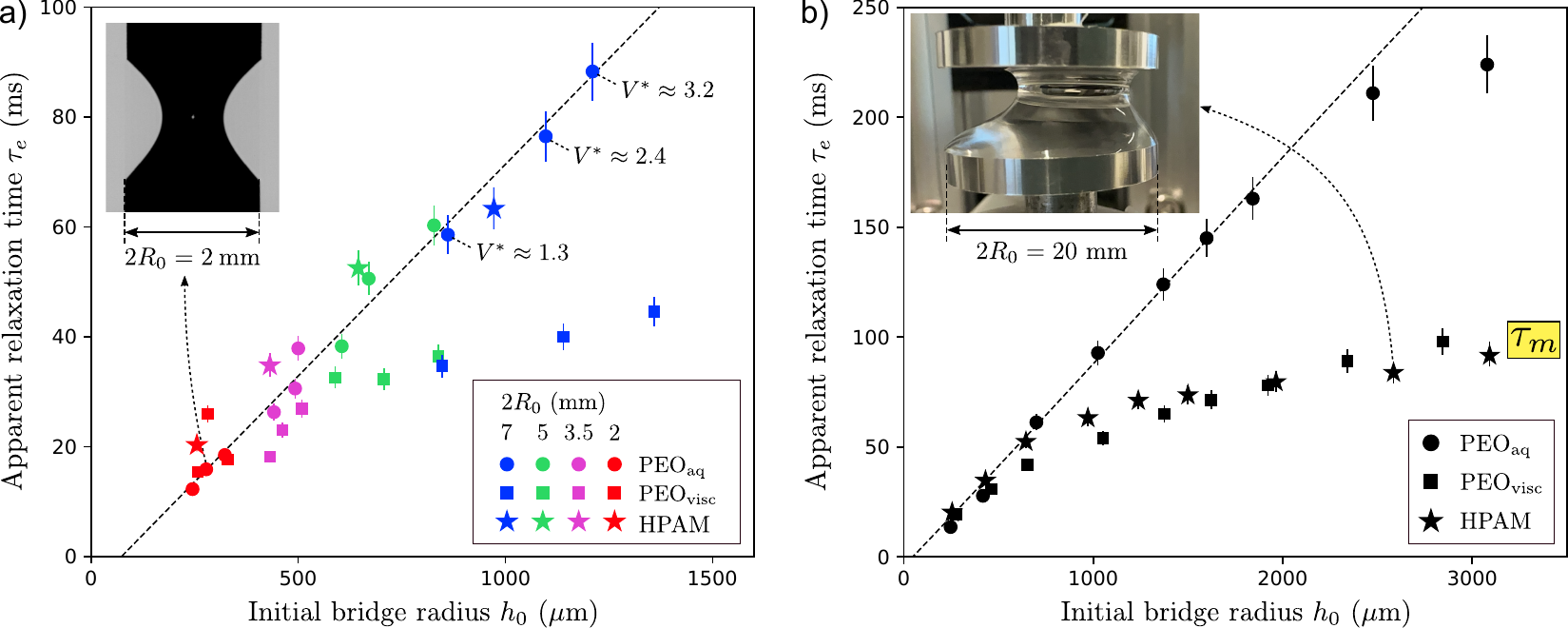}}
  \caption{Apparent relaxation time $\tau_e$ against $h_0$ for all three solutions for plate diameters $2R_0 = 2$, $3.5$, $5$ and $7$~mm (a) and for $2R_0 = 2$, $3.5$, $5$, $7$, $10$, $12.5$, $15$, $20$ and $25$~mm (b). In (a), data points of the same colour correspond to the same $R_0$ with different droplet volumes $V^* = V/R_0^3 \approx 1.3$, $2.4$ and $3.2$ (only $V^* \approx 2.4$ for the HPAM solution). In (b), a single volume is used for each plate diameter ($V^* \approx 2.4$ for the smallest plates and $0.88$ for the largest plates). The inset images show stable liquid bridges for $2R_0 = 2$~mm (a) and $20$~mm (b). The linear fit is for the PEO$_{\mathrm{aq}}$ solution for $h_0 < 2$~mm.}
\label{fig:taue}
\end{figure}

The apparent relaxation time $\tau_e$ measured with our slow stepwise plate separation CaBER protocol is plotted in figure \ref{fig:taue}(a) as a function of the initial bridge radius $h_0$ for various plate diameters and droplet volumes for all three polymer solutions, data points of the same colour corresponding to the same $R_0$ with different $V^*$. We observe that $\tau_e$ increases with both $R_0$ and $V^*$ and that all data points for a given solution collapse on a single curve when plotted against $h_0$, which is itself an increasing function of both $R_0$ and $V^*$. In other words, a given solution tested with two different ($R_0,V^*$) sets but with the same $h_0$ yields the same $\tau_e$, as some examples show in figure \ref{fig:taue}(a). This suggests that $h_0$ is in fact the only relevant geometrical parameter of the problem. This is in agreement with the accepted idea that polymer deformations during capillary thinning are only influenced by the local extensional flow in the bridge / filament of maximum extension rate $\dot{\epsilon} = - 2 \dot{h}/h$ at its thinnest point, while the top and bottom end droplets act as passive liquid reservoirs, their size not directly influencing the pinch-off dynamics.

The apparent relaxation time varies significantly (up to a factor $4$) within the typical range of plate diameters used for CaBER experiments, see figure \ref{fig:taue}(a). However, $\tau_e$ cannot increase indefinitely with increasing $h_0$. In order to observe the expected saturation of $\tau_e$ for larger $h_0$ values, we had to move to much larger plate diameters, up to $2R_0 = 25$~mm. For plate diameters $2R_0 \ge 10$~mm, the top end-drop does not cover the top plate fully because of gravity, as shown in the inset image of figure \ref{fig:taue}(b). In fact, there is always a thin liquid film covering the top plate due to the plasma treatment. For such large plates, the top end-drop is not at the centre of the the top plate since the two plates are not perfectly parallel.

In spite of this lack of full coverage for large plates, we find that the critical minimum bridge radius $h_0$ marking the onset of the Rayleigh-Plateau instability increases with $R_0$, allowing us to explore a wider range $h_0$ values, as shown in figure \ref{fig:taue}(b) where the apparent relaxation time $\tau_e$ seems to saturate to a maximum value $\tau_m$, reported in table \ref{tab:rheology_shortpaper}, at large $h_0$. Since no clear plateau is observed, especially for the PEO$_{\mathrm{aq}}$ solution, the value of $\tau_m$ is only an estimation. In figure \ref{fig:taue}(b), we only show one data point for each plate diameter, with  $V^*$ between $2.4$ for the smallest plates and $0.88$ for the largest plates. Note that no change in behaviour is observed in the $\tau_e(h_0)$ curves in figure \ref{fig:taue}(b) at the transition between fully covered ($2R_0 \le 7$~mm) and not fully covered top plates ($2R_0 \ge 10$~mm), around $h_0 \approx 1.3$~mm, strengthening the claim that the top and bottom end-drops are passive liquid reservoirs whose size and shape do not affect the filament thinning dynamics.

As shown by the inset images in figure \ref{fig:taue}(a,b), the bottom end-drop becomes increasingly larger than the top one as $R_0$ increases since the Bond number $\Bo = \rho g R_0^2 / \gamma$ ranges between $0.16$ and $25$. However, the thinning dynamics is not driven by gravity since the ``filament'' Bond number $\Bo_f = \rho g L_f h_1 / \gamma$, comparing the typical capillary pressure $\gamma / h_1$ in the filament to the hydrostatic pressure $\rho g L_f$ over the filament length $L_f$, is only up to $0.1$ for the largest plates. This is also evident from the fact that filaments are not thicker at their base, see, e.g., the right inset image in figure \ref{fig:protocol}(a). This is consistent with the fact that, as we discuss in our longer companion paper \cite{gaillard2024does}, the bridge thinning dynamics is well captured by the classical inertio-capillary and visco-capillary (gravity-free) self-similar laws (close to the transition to the elastic regime) for the PEO$_{\mathrm{aq}}$ and PEO$_{\mathrm{visco}}$ solutions, respectively, for plate diameters where $\tau_e$ depends on $h_0$ (hence suggesting that the $h_0$-dependence of $\tau_e$ is not caused by gravitational drainage). These apparent extensional relaxation times $\tau_e$ are compared with relaxation times inferred from shear rheology in our companion paper \cite{gaillard2024does}.

Note that the PEO solutions used in figures \ref{fig:taue}(b) are not the same as the ones used in figure \ref{fig:taue}(a) and have apparent relaxation times about $30$\% larger for $2R_0 = 7$~mm, while the shear viscosity was only up to $10$~\% larger, meaning that the shear rheology parameters in table \ref{tab:rheology_shortpaper} are representative of both solutions. These differences are due to slightly different preparation protocols, e.g. agitation times, for a given recipe. These extra solutions were prepared because, by the time we had realised much larger plates were needed to observe the saturation of $\tau_e$, the previous solutions had considerably aged, i.e., had lower $\tau_e$ values.

\section{Interpretations}
\label{sec:Interpretations}

\noindent The apparent disagreement between experiments and equation \ref{eq:exponential} implies that either the liquid changes, becoming less elastic for lower values of $h_0$, or that the Oldroyd-B model, from which equation \ref{eq:exponential} is derived, misses some important features of polymer dynamics in extensional flows. We now consider some possible explanations.

\subsection{Evaporation and degradation}
\label{subsec:Interpretations0}

First, although the relaxation time measured in filament thinning is known to increase with polymer concentration \cite{tirtaatmadja2006drop,clasen2006dilute}, solvent evaporation cannot explain the observed increase of the apparent relaxation time with increasing droplet size. Indeed, the bulk polymer concentration would increase quicker for smaller droplets due to their larger surface to volume ratio, leading to larger concentrations for smaller droplets, and hence larger $\tau_e$, while the opposite is observed (lower $\tau_e$ for lower $h_0$). Besides, repeating an experiment several times over the course of $10$ minutes does not lead to a monotonic increase or decrease of $\tau_e$ over time, beyond small variations of less than $5$\%. This is not surprising since humidity is kept at high levels ($>80$\%) in the CaBER chamber. The latter observation also argues against polymer degradation as a possible explanation. Moreover, $\tau_e$ is observed to increase with $h_0$ for both PEO and HPAM solutions, even though HPAM is less fragile than PEO.  

Therefore, if the liquid is in fact the same for each experiment, the Oldroyd-B model fails to describe the full polymer dynamics in the bridge / filament. In particular, differences in the history of polymer deformations for different drop sizes could lead to different ``initial'' states of polymers at the onset of the elastic regime, which could result in different filament thinning rates. We now discuss whether finite extensibility of polymer chains, as described by the FENE-P model, can account for such differences. 

\subsection{Elasto-capillary balance with FENE-P}
\label{subsec:Interpretations1}

Following Wagner et al. \cite{wagner2015analytic}, for a uniaxial extensional flow, the polymeric part of the normal stress is $\sigma_{p,zz} = G (f A_{zz} - 1)$ in the flow direction $z$ where $G$ is the elastic modulus and $A_{zz}$ is the normal part of the conformation tensor ${\bf A}$ which follows
\begin{equation}
\dot{A}_{zz} - 2 \dot{\epsilon} A_{zz} = - \frac{f A_{zz}-1}{\tau},
\label{eq:FENEP_0}
\end{equation}
\noindent where $\dot{\epsilon}$ is the extension rate, $\tau$ the relaxation time and $f = (1 - \mathrm{tr}({\bf A})/L^2)^{-1}$ where $L$ is the ratio of the fully unravelled chain size to its equilibrium size. In this model the stress diverges as $A_{zz}$ approaches its limit value $L^2$. 

In the elastic regime ($t \ge t_1$), we assume that polymers are far from equilibrium and that the axial stress dominates over the radial stress, i.e. $A_{zz} \gg 1 > A_{rr}$. Assuming negligible inertia and solvent viscosity in the elastic regime, we use the elasto-capillary force balance equation 
\begin{equation}
(2X-1) \frac{\gamma}{h} = \sigma_{p,zz} = G f A_{zz} \,\mathrm{,}
\label{eq:FENEP_1}
\end{equation}
\noindent with $f = (1 - A_{zz}/L^2)^{-1}$ where $X=3/2$ in the Oldroyd-B limit \cite{eggers2020self}. Assuming a small correction due to finite extensibility, combining equations \ref{eq:FENEP_0} and \ref{eq:FENEP_1} with $\dot{\epsilon} = -2 \dot{h}/h$ leads to the ordinary differential equation $(3+A_{zz}/L^2) \dot{A}_{zz} = A_{zz}/\tau$ which has an implicit solution
\begin{equation}
\frac{t-t_1}{\tau} = \frac{A_{zz}-A_1}{L^2} + 3 \ln{\left( \frac{A_{zz}}{A_1} \right)} \,\mathrm{,}
\label{eq:FENEP_5}
\end{equation}
\noindent where $A_1 = A_{zz}(t_1)$ quantifies the amount of polymer stretching at the onset of the elastic regime at time $t_1$. The filament radius can be computed by noticing that $h f A_{zz}$ is a constant according to equation \ref{eq:FENEP_1}, i.e.
\begin{equation}
\frac{h}{h_1} = \frac{f_1 A_1}{f A_{zz}} \,\mathrm{,}
\label{eq:FENEP_6}
\end{equation}
\noindent where $f_1 = (1-A_1/L^2)^{-1}$ and $h_1 = h(t_1)$ is the filament radius at the onset of the elastic regime. $h$ depends on three parameters: $\tau$, $h_1$ and the ratio $A_1/L^2$ quantifying how far chains are from being fully extended at the onset of the elastic regime. Indeed, according to equations \ref{eq:FENEP_5} and \ref{eq:FENEP_6}, $h$ is unchanged upon multiplying both $A_1$ and $L^2$ by the same quantity. In the Oldroyd-B limit $L^2 \to \infty$ ($f=1$), we recover the expected exponential trends $A_{zz} = A_{1} \exp{((t-t_1)/3\tau)}$ and $h = h_1 \exp{(-(t-t_1)/3\tau)}$. For a finite extensibility, the exponential regime holds until $A_{zz}$ approaches $L^2$ where finite extensibility effects arise. Ultimately, the stress diverges and $h \to 0$ in finite time when $A_{zz}$ saturates to $L^2$, which occurs sooner as $A_1/L^2$ is closer to one. In particular, if $A_1/L^2$ is only slightly less than $1$, meaning that polymer chains are already almost fully extended at the onset of the elastic regime, finite extensibility effects are never negligible and equation \ref{eq:exponential} is never valid. In that case, increasingly larger filament thinning rates are be observed as $A_1/L^2$ increases and equations \ref{eq:FENEP_5} and \ref{eq:FENEP_6} predict that the apparent relaxation time $\tau_e$ is well approximated by $\tau_e / \tau \approx 1 - A_1/L^2$.

\begin{figure}[ht!]
  \centerline{\includegraphics[scale=0.57]{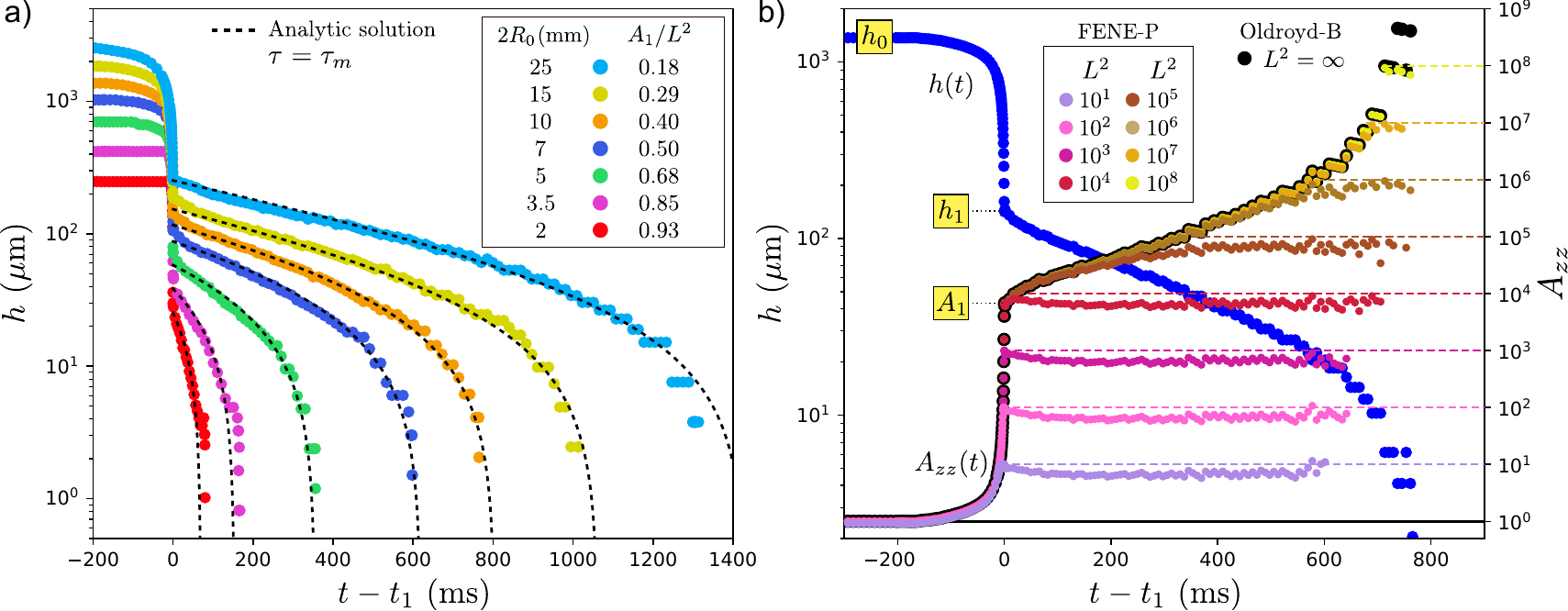}}
  \caption{(a) Experimental $h(t)$ for the PEO$_{\mathrm{aq}}$ solution tested with different plate diameters. The elastic regime ($t \ge t_1$) is fitted by equation \ref{eq:FENEP_5} and \ref{eq:FENEP_6} with relaxation time $\tau = \tau_m$ (from experiments, table \ref{tab:rheology_shortpaper}), using $h_1$ and $A_1/L^2$ as fitting parameters. Fitting values of the parameter $A_1/L^2$ are given in the legend. (b) Experimental $h(t)$ for the PEO$_{\mathrm{aq}}$ solution tested with a single plate diameter $2R_0 = 10$~mm and corresponding values of $A_{zz}(t)$ calculated from equation \ref{eq:FENEP_0} with $\dot{\epsilon} = -2 \dot{h}/h$ using the experimental values of $h$, with relaxation time $\tau = \tau_m$ (from experiments, table \ref{tab:rheology_shortpaper}) where we vary the value of $L^2$ from $10$ to $10^8$ (FENE-P) and $L^2 = \infty$ (Oldroyd-B), using $A_{zz}=1$ at $h=h_0$ as the initial condition. The Ohnesorge numbers $\Oh = \eta_0 / \sqrt{\rho \gamma h_0}$ ranges between $0.02$ and $0.007$ for the PEO$_{\mathrm{aq}}$ solution in our range of $h_0$ values which corresponds to plates diameters $2R_0$ between $2$ and $25$~mm.}
\label{fig:h_Azz}
\end{figure}

This theory is tested in figure \ref{fig:h_Azz}(a) where the elastic regime ($t \ge t_1$) of the PEO$_{\mathrm{aq}}$ solution, tested with different plate diameters, is compared with the predictions of equations \ref{eq:FENEP_5} and \ref{eq:FENEP_6} where we have chosen the maximum relaxation time $\tau_m$ (table \ref{tab:rheology_shortpaper}) measured at large $h_0$ as the relaxation time of the FENE-P model. We use $A_1/L^2$ and $h_1$ as fitting parameters to obtain a good agreement between model and experiments. Most importantly, we have to impose that $A_1/L^2$ gets closer to one as $h_0$ decreases to capture the observed thinning rates, all larger than $1/3 \tau_m$. For $2R_0 = 2$~mm for example, we need $A_1/L^2 = 0.93$, meaning that polymers are almost fully extended at the onset of the elastic regime. 

We emphasise here that in previous studies, where $\tau_e$ was believed to not vary with the plate size, comparisons with the FENE-P model were performed using $\tau_e$ as the model's relaxation time, and it is quite remarkable that when using a larger value $\tau_m$, one can still obtain a somehow exponential-looking trend with the right thinning rate by tuning $A_1/L^2$ for $\tau_e < \tau_m$. Equally good fits can be obtained for the PEO$_{\mathrm{visc}}$ and HPAM solutions and the corresponding values of the fitting parameter $A_1/L^2$ are plotted against $h_0$ in figure \ref{fig:A1}(a) (light purple). These results suggest that the maximum relaxation time $\tau_m$ measured at large plate sizes could be the `true' relaxation time, lower apparent values $\tau_e$ being a consequence of polymers being too close to their finite extensibility limit at the onset of the elastic regime to display their `natural' (far from full extension) relaxation behaviour.

\subsection{Calculating $A_{zz}(t)$ from experimental $h(t)$ with FENE-P}
\label{subsec:Interpretations2}

In \S\ref{subsec:Interpretations1}, we used $A_1/L^2$ as a fitting parameter and showed that it should increase towards $1$ as $h_0$ decreases in order to successfully describe the elastic regime. However, values of $A_1$ can in fact be calculated from the flow history in the Newtonian regime ($t<t_1$). Therefore, we now check the consistency of the proposed interpretation for the variation of $\tau_e$ with $h_0$ by calculating $A_1$ from the FENE-P model, examining whether polymers are indeed expected to be more stretched at the onset of the elastic regime for smaller plate diameters or not.

To this end, we use equation \ref{eq:FENEP_0} to calculate $A_{zz}(t)$, using the experimental values of $h(t)$ for $\dot{\epsilon}(t) = -2 \dot{h}/h$, although this expression is only valid at the thinnest bridge radius. In other words, we calculate the predictions of the model for the experimental history of extension rates. In particular, the extension rate history in the Newtonian regime ($t<t_1$) sets $A_1$. Hence, we do \emph{not} assume large polymer deformations ($A_{zz} \not\gg 1$) since, as our slow stepwise plate separation protocol is designed for, polymers are at equilibrium at the onset of capillary thinning, i.e. $A_{zz} = 1$ when $h=h_0$. We use $f = (1-A_{zz}/L^2)^{-1}$ since when $f$ is not close to $1$ anymore, the axial stress dominates over radial stress. In order to circumvent the issue of calculating $\dot{h}$ from experimental values of $h$, we introduce a function $y(t)$ such that $A_{zz} = y / h^4$, which gives $\dot{A}_{zz} + 4 (\dot{h}/h) A_{zz} = \dot{y}/h^4$, so that equation \ref{eq:FENEP_0} becomes $\tau \dot{y} = h^4 - y / (1 - y/ (h^4 L^2))$ which does not involve $\dot{h}$ anymore. To solve this equation, we use a standard ODE solver, using spline interpolation to create a $t \to h(t)$ function based on experimental values of $h$. This equation can be integrated analytically in the Oldroyd-B limit, as shown by Bazilevsky et al. \cite{bazilevsky2001breakup}.

The results are shown in figure \ref{fig:h_Azz}(b) for the PEO$_{\mathrm{aq}}$ solution tested with a plate diameter $2R_0 = 10$~mm, with $\tau = \tau_m$ (table \ref{tab:rheology_shortpaper}) for the relaxation time of the FENE-P model, along with various values of $L^2$, including the Oldroyd-B limit $L^2 \to + \infty$. As expected, values of $A_{zz}$ calculated from FENE-P coincide with Oldroyd-B until it saturates when reaching $L^2$. In particular, the value of $A_1$ becomes independent of $L^2$ when $L^2$ is sufficiently large and becomes indistinguishable from the values predicted by the Oldroyd-B model.

\subsection{Comparing fitting and calculated $A_1$ values}
\label{subsec:Interpretations3}

\begin{figure}[t]
  \centerline{\includegraphics[scale=0.57]{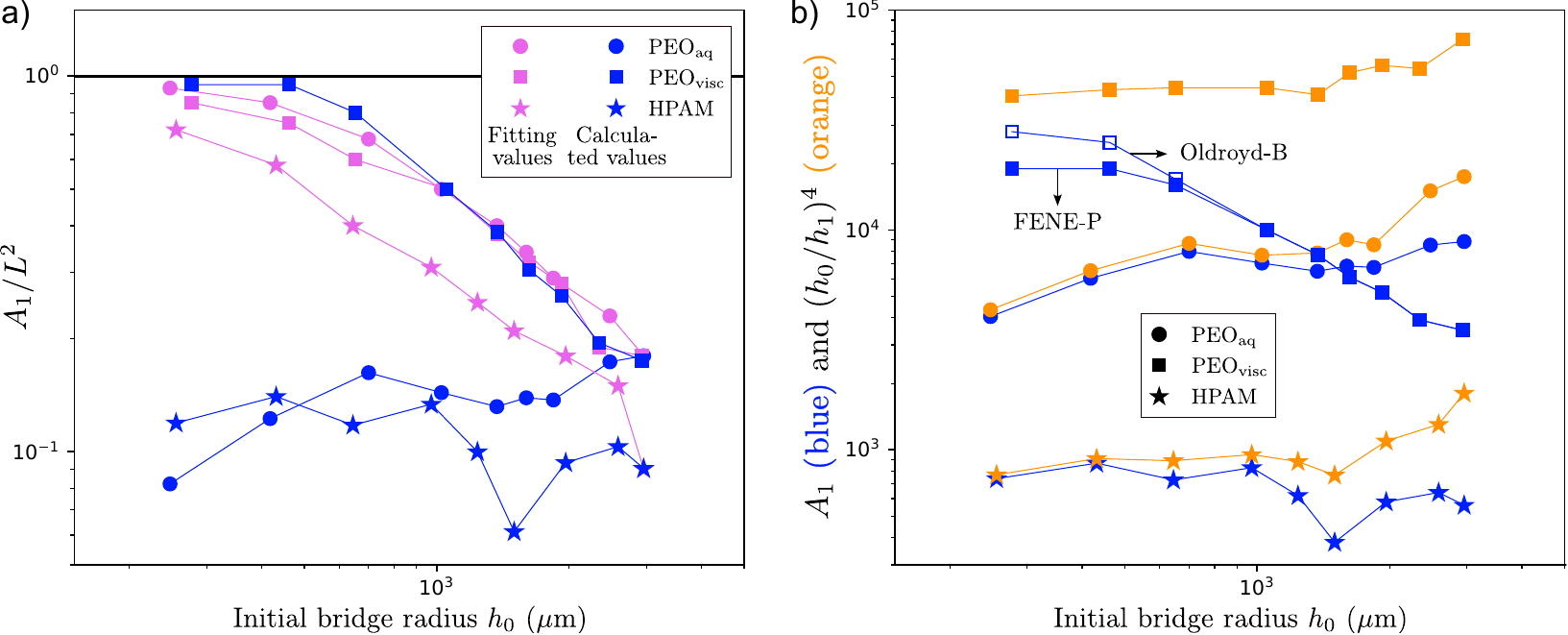}}
  \caption{(a) Values of $A_1/L^2$ used as fitting parameters (light purple, see e.g. figure \ref{fig:h_Azz}(a)), and values calculated from the FENE-P model (dark blue, see e.g. figure \ref{fig:h_Azz}(b)) for $\tau = \tau_m$ and values of $L^2$ discussed in the text, against $h_0$ for all liquids and plate diameters. (b) Same values of $A_1$ calculated from the FENE-P model (dark blue), compared with $(h_0/h_1)^4$ (light orange). All values of $A_1$ calculated from FENE-P are the same as in Oldroyd-B except for the PEO$_{\mathrm{visc}}$ solution at low $h_0$ where Oldroyd-B values are shown with empty blue symbols. The Ohnesorge numbers $\Oh = \eta_0 / \sqrt{\rho \gamma h_0}$ ranges between $0.02$ and $0.007$ for the PEO$_{\mathrm{aq}}$ solution, between $0.1$ and $0.03$ for the HPAM solution and between $2$ and $0.6$ for the PEO$_{\mathrm{visc}}$ solution in our range of $h_0$ values which corresponds to plates diameters $2R_0$ between $2$ and $25$~mm.}
\label{fig:A1}
\end{figure}

The values of $A_1 = A_{zz}(t_1)$ calculated in \S\ref{subsec:Interpretations2} from the experimental values of $h(t)$ in the Newtonian regime ($t<t_1$), using the FENE-P model with $\tau = \tau_m$, are plotted in figure \ref{fig:A1}(a) against $h_0$ for all three solutions (dark blue). More precisely, we plot the ratio $A_1/L^2$ indicating how close polymer chains are to being fully extended at the onset of the elastic regime (full extension corresponding to $A_1/L^2 = 1$). This is because we want to compare these calculated values (dark blue) with the values of $A_1/L^2$ used as a fitting parameter (light purple) in \S\ref{subsec:Interpretations1} to fit the elastic regime with equations \ref{eq:FENEP_5} and \ref{eq:FENEP_6} (see figure \ref{fig:h_Azz}(a)). We hence need to choose a value of $L^2$. For each liquid, we choose $L^2$ such that, at the largest $h_0$, the calculated value (dark blue) of $A_1/L^2$ coincides exactly with the fitting value (light purple). We obtain $L^2 = 4.9 \times 10^{4}$ for the PEO$_{\mathrm{aq}}$ solution, $L^2 = 2.0 \times 10^{4}$ for the PEO$_{\mathrm{visc}}$ solution and $L^2 = 6.2 \times 10^{3}$ for the HPAM solution. The order of magnitude is consistent with the microscopic formula \cite{clasen2006dilute}
\begin{equation}
L^2 = 3 \left[ \frac{j \sin^2{(\theta/2)} M_w}{C_{\infty} \, M_u} \right]^{2(1-\nu)} \,\mathrm{,}
\label{eq:L2_microscopic}
\end{equation}
\noindent which gives $L^2$ between $4.5 \times 10^{4}$ and $1.3 \times 10^{5}$ for PEO of molecular weight $M_w = 4 \times 10^{6}$~g/mol, for typical solvent quality exponents $\nu$ between $0.55$ and $0.5$ (theta solvent) found for PEO in water-based solvents, where $M_u$ is the monomer molecular weight, $\theta = 109^{\circ}$ the C-C bond angle, $j=3$ the number of bonds of a monomer and $C_{\infty} = 4.1$ the characteristic ratio \citep{brandrup1989polymer}. We find that, while the fitting values of $A_1$ (light purple) increase towards $L^2$ as $h_0$ decreases, the calculated values of $A_1$ (dark blue) only do so for the PEO$_{\mathrm{visc}}$ solution, for which a good agreement is found with the fitting values, and do not for the PEO$_{\mathrm{aq}}$ and HPAM solutions for which $A_1$ is fairly constant. For these last two, no other value of $L^2$ can lead to a better agreement since decreasing $L^2$ would just shift all calculated values towards the upper limit $A_1/L^2=1$.

In order to better understand this, we compare these calculated values of $A_1$ (dark blue) with their upper limit $(h_0/h_1)^4$ (light orange) in figure \ref{fig:A1}(b). This upper limit corresponds to a relaxation time $\tau$ so large that polymer relaxation (right hand side of equation \ref{eq:FENEP_0}) is always negligible in the Newtonian regime ($t<t_1$), a case where equation \ref{eq:FENEP_0} (with $\dot{\epsilon} = -2 \dot{h}/h$) can be integrated as $A_{zz} h^4 = h_0^4$. Differences in values of $(h_0/h_1)^4$ among the three polymer solutions are due to differences in $h_1$ stemming from different elastic moduli $G$, as we discuss in our longer companion paper \cite{gaillard2024does} where we show that the Oldroyd-B model gives $h_1 \propto (G H^4 / \gamma)^{1/3}$ where $H \to h_0$ for large relaxation times. We find in figure \ref{fig:A1}(b) that $A_1$ is very close to the $(h_0/h_1)^4$ limit for the PEO$_{\mathrm{aq}}$ and HPAM solutions at low $h_0$, meaning that polymer relaxation is indeed negligible, and that the ratio between the two increases as $h_0$ increases, meaning that relaxation becomes more important. This is consistent with the fact that the Deborah number $\Deb = \tau_m / \tau_R$ decreases from $500$ ($\gg 1$, i.e., negligible relaxation) at the lowest $h_0$ to $5$ at the highest $h_0$, where $\tau_R = \sqrt{\rho h_0^3 / \gamma}$ is the Rayleigh time scale relevant for the thinning dynamics of such low-viscosity liquids, with Ohnesorge numbers $\Oh = \eta_0 / \sqrt{\rho \gamma h_0}$ which are up to $0.02$ for PEO$_{\mathrm{aq}}$ and up to $0.1$ for HPAM (at the lowest $h_0$ values), i.e. $\Oh \ll 1$. The reason why the calculated values of $A_1$ are fairly independent of $h_0$ for these two liquids, hence resulting in an impossible match with the fitting values of $A_1/L^2$ (light purple) in figure \ref{fig:A1}(a), is therefore because these calculated values of $A_1$ are close to their upper limit $(h_0/h_1)^4$ which are fairly independent of $h_0$ themselves (at least for sufficiently low $h_0$, see figure \ref{fig:A1}(b)). Note that the scaling $h_1 \propto h_0$ implied by the constant values of $(h_0/h_1)^4$ is discussed in our longer companion paper \cite{gaillard2024does}.

By contrast, the calculated values of $A_1$ for the PEO$_{\mathrm{visc}}$ solution are increasing as $h_0$ decreases, allowing for a good match with the fitting values of $A_1/L^2$ (light purple) in figure \ref{fig:A1}(a). This is because, unlike for the two other solutions, polymer relaxation (right hand side of equation \ref{eq:FENEP_0}) is not negligible in the Newtonian regime ($t<t_1$), as indicated by the greater difference between $A_1$ and the upper (relaxation-free) limit $(h_0/h_1)^4$ for the PEO$_{\mathrm{visc}}$ solution in figure \ref{fig:A1}(b). This is due to the slower thinning dynamics in the Newtonian regime, caused by a larger shear viscosity with Ohnesorge numbers $\Oh = \eta_0 / \sqrt{\rho \gamma h_0}$ ranging between $0.6$ and $2$ in our range of $h_0$ values. Indeed, since all three solutions have comparable relaxation times, slower thinning dynamics mean that polymer relaxation is more important, i.e., $A_1 < (h_0/h_1)^4$. The reason why $A_1$ decreases as $h_0$ increases for the PEO$_{\mathrm{visc}}$ solution is because the ratio between $A_1$ and $(h_0/h_1)^4$ increases with $h_0$. This is because relaxation is increasingly important as $h_0$ increases since the time scale of the Newtonian thinning dynamics, expected to scale as $\tau_R \propto h_0^{3/2}$ or as $\tau_{\mathrm{visc}} = \eta_0 h_0/\gamma \propto h_0$ depending on $\Oh$, increases with $h_0$, resulting in lower Deborah numbers (based on either $\tau_R$ or $\tau_{\mathrm{visc}}$).

\subsection{Numerical simulations}
\label{subsec:Interpretations4}

\begin{figure}[t!]
  \centerline{\includegraphics[scale=0.57]{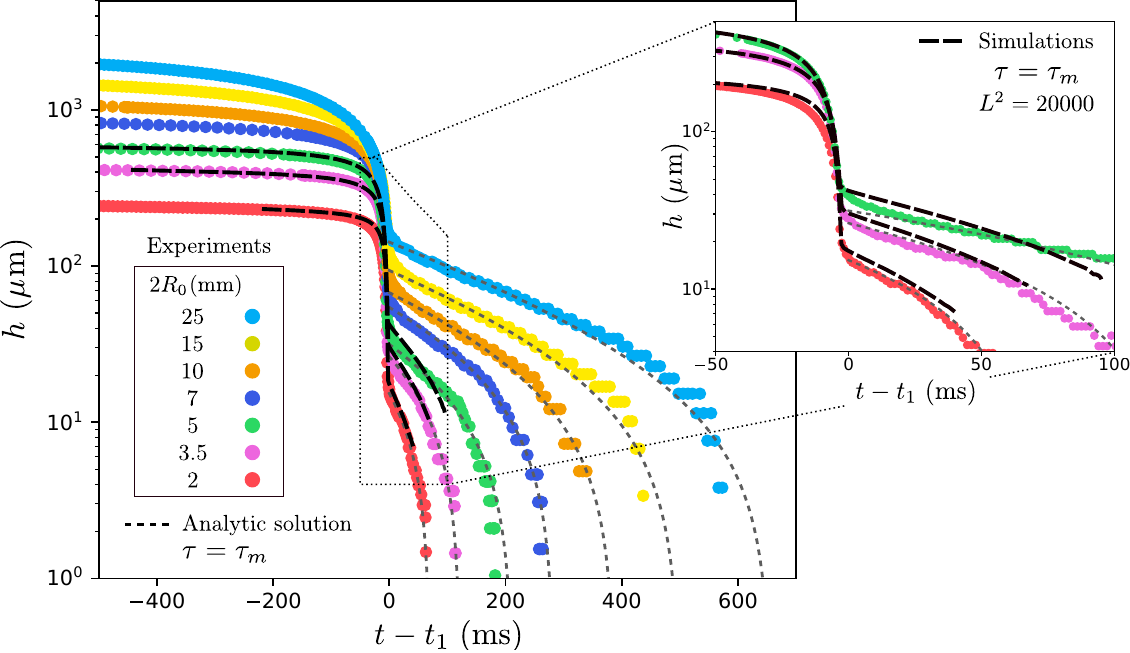}}
  \caption{$h(t)$ from experiments and simulations for the PEO$_{\mathrm{visc}}$ solution for different plate diameters. Simulations are performed for $2R_0 = 2$, $3.5$ and $5$~mm only, using the FENE-P model with $\tau = \tau_m$ and $L^2 = 2.0 \times 10^4$. The elastic regime is fitted by equations \ref{eq:FENEP_5} and \ref{eq:FENEP_6} with $\tau = \tau_m$ where, like in figure \ref{fig:h_Azz}(a), $A_1/L^2$ and $h_1$ are used as fitting parameters. The Ohnesorge numbers $\Oh = \eta_0 / \sqrt{\rho \gamma h_0}$ ranges between $2$ and $0.6$ for the PEO$_{\mathrm{visc}}$ solution in our range of $h_0$ values which corresponds to plates diameters $2R_0$ between $2$ and $25$~mm.}
\label{fig:simu}
\end{figure}

Hence, The FENE-P model can only explain the increase of the apparent relaxation time $\tau_e$ with $h_0$ for the PEO$_{\mathrm{visc}}$ solution which is the most dilute one and with the highest solvent viscosity out of our three solutions. This is checked further by numerical simulations of the axisymmetric problem (with gravity) using the full FENE-P constitutive equation. We use as fixed model parameters the values of $\eta_s$ and $\eta_p$ from the shear rheology, the high-$h_0$ limit $\tau = \tau_m$ for the relaxation time and the value of $L^2 = 2.0 \times 10^{4}$ used in \S\ref{subsec:Interpretations3} (figure \ref{fig:A1}(a)) for which the simplified analytical model in \S\ref{subsec:Interpretations1} and \S\ref{subsec:Interpretations2} could rationalise the apparent relaxation times $\tau_e$. The equations to be solved are the same as in Rubio et al  \cite{rubio2020breakup} and the numerical methods are detailed in \ref{appC}. The initial condition is established by starting from a stable liquid bridge with a plate-to-plate distance $L_p$ just below the instability threshold value and slightly increasing $L_p$ to trigger the pinch-off.

The results are shown in figure \ref{fig:simu} for the three smallest plates and the corresponding experimental droplet volumes in terms of the time evolution of the minimum bridge / filament radius. Simulations are found to start at a bridge radius close to $h_0$, which validates the numerical method to set the initial condition. We find that simulations are able to capture the Newtonian regime quite well and provide a reasonable agreement with experiments in the elastic regime. In particular, the filament thinning rate varies with the plate diameter, consistent with experiments, while the Oldroyd-B model would give the same (constant) thinning rate $1/3\tau_m$. Simulations could not be continued far enough to compare with the full experimental time window. Like in figure \ref{fig:h_Azz}(a), figure \ref{fig:simu} also features the analytic solution of equations \ref{eq:FENEP_5} and \ref{eq:FENEP_6} for the elastic regime ($t>t_1$).

\section{Conclusions and discussion}
\label{sec:Conclusions and discussions}

\noindent We have shown experimentally that the thinning rate of filaments of various polymer solutions is not necessarily just a material property but may depend on the size of the system in CaBER with both slow (stepwise) and fast (step-strain) plate separation protocols as well as in DoS experiments, consistent with previous observations for dripping experiments \cite{rajesh2022transition}. Although all filaments are observed to thin exponentially, as predicted by the Oldroyd-B model (see equation \ref{eq:exponential}), we show that, for CaBER with slow stepwise plate separation, the inferred apparent relaxation time $\tau_e$ increases with the minimum bridge radius $h_0$ marking the onset of capillary thinning, which is an increasing function of both the plate diameter and droplet volume, and that $\tau_e$ saturates at large $h_0$ values corresponding to plate diameters $>10$~mm significantly larger than typical CaBER plates. These observations hence suggest that CaBER relaxation times reported in the literature are not universal since testing a given polymer solution with different plate diameters and droplet volumes can yield significantly different results.

The fact that Bazilevsky et al. \cite{bazilevsky1997failure}, who used both fast and slow-retraction CaBER methods, reported no variation of $\tau_e$ with the drop volume $V$ (without providing the data to support their claim) might be due to the fact that its dependence on $V$ is weak (weaker than its dependence on $R_0$, see figure \ref{fig:taue}(a)) and that, for a given plate diameter, $V$ can only be varied up to a critical value above which the drop does not fit on the plate.

We demonstrate that the variation of $\tau_e$ with $h_0$ is not caused by solvent evaporation or polymer degradation and cannot be universally explained by finite extensibility effects described by the FENE-P model. These observations suggest that the single-mode Oldroyd-B and FENE-P models miss some important features of polymer dynamics in extensional flows. The FENE-P model could only explain the variation of $\tau_e$ for the most dilute solution with the most viscous solvent, which is consistent with the fact that (i) the FENE-P model is derived for dilute solutions and that (ii) inertio-capillary oscillations are absent for this solution. However, since the value of the finite-extensibility parameter $L^2$ was chosen to optimise the agreement with experiments, we do not exclude that this agreement may also be a coincidence, although this value in agrees with the microscopic prediction (equation \ref{eq:L2_microscopic}).

A physical interpretation for this deformation-history-dependent filament thinning rate is still needed, strengthening the already established need for better constitutive equations. Other shortcomings of the FENE-P model, such as coil-stretch hysteresis and the increase of $\tau_e$ with the polymer concentration in the `dilute' regime ($c < c^*$), were previously explained by a Conformation-Dependent Drag (CDD) model accounting for the action of both chain stretching and intermolecular hydrodynamic interactions on the friction coefficient \cite{prabhakar2016influence,prabhakar2017effect}. Future works will determine if such models are also able to capture the system-size dependence of the effective relaxation time discussed in this paper.\\

\noindent \textbf{Declaration of Interests}. The authors report no conflict of interest.

\textbf{Funding} M. A. Herrada acknowledges funding from the Spanish Ministry of Economy, Industry and Competitiveness under Grant PID2022-140951O.

\noindent \textbf{Acknowledgements}. We thank Louison Laruelle and Carmen van Poelgeest for preliminary experimental work. \\

\appendix
\section{Dripping-onto-substrate (DoS)}
\label{appA}

\begin{figure}[t!]
  \centerline{\includegraphics[scale=0.57]{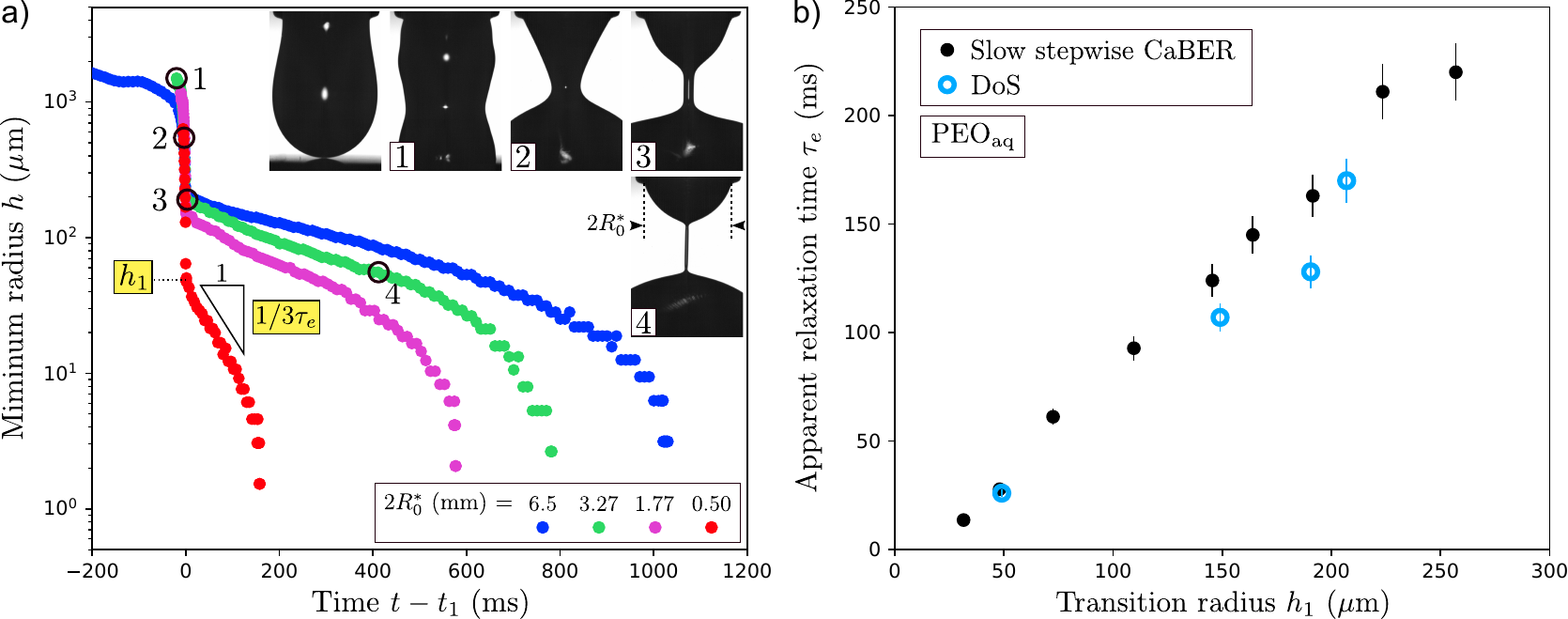}}
  \caption{(a) $h(t)$ from DoS experiments with different nozzle diameters for the PEO$_{\mathrm{aq}}$ solution. We report the values of the maximum diameter $2R_0^*$ of the top end-drop which is between the inner and outer nozzle diameter. Insert images correspond to the steadily hanging drop (left) and to four times labelled 1 to 4 indicated on the $h(t)$ curve for $2R_0^* = 3.27$~mm. (b) $\tau_e$ vs. $h_1$ for the PEO$_{\mathrm{aq}}$ solution from DoS compared to the values of figure \ref{fig:taue}(b) from CaBER with the slow stepwise plate separation protocol described in \S\ref{subsec:Slow stepwise plate separation CaBER protocol}.}
\label{fig:DoS}
\end{figure}

\noindent In DoS experiments, a horizontal substrate (here, a plasma-treated aluminium plate) is moved slowly upward until being in contact with a liquid droplet hanging steadily from a nozzle. As shown in the image sequence in figure \ref{fig:DoS}(a), a fast spreading of the liquid on the plate leads to the pinch-off of the bridge connecting the substrate to the nozzle. This transiently leads to the formation of an exponentially thinning filament, as shown by the time-evolution of the minimum bridge / filament radius $h$ in figure \ref{fig:DoS}(a), where the PEO$_{\mathrm{aq}}$ solution is tested with four different nozzle diameters. As in CaBER experiments, the apparent relaxation time extracted from the filament thinning rate increases with the droplet size, here quantified by the nozzle diameter. This apparent relaxation time $\tau_e$ is plotted in figure \ref{fig:DoS}(b) for both CaBER and DoS experiments against the filament radius $h_1$ marking the onset of the elastic regime which, unlike $h_0$ in CaBER, is easily definable in both methods. The relatively good collapse of the data points on a single curve suggests a universal physical mechanism for the dependence of the apparent relaxation time on the size of the system, independent of the exact method used. We checked that $\tau_e$ also increases with the nozzle diameter when the droplet spreads on a `small' plate (about two times larger than the nozzle and made of non-plasma-treated aluminium), where spreading stops before the viscoelastic filament is formed.

\section{Step-strain CaBER}
\label{appB}

\begin{figure}[t!]
  \centerline{\includegraphics[scale=0.57]{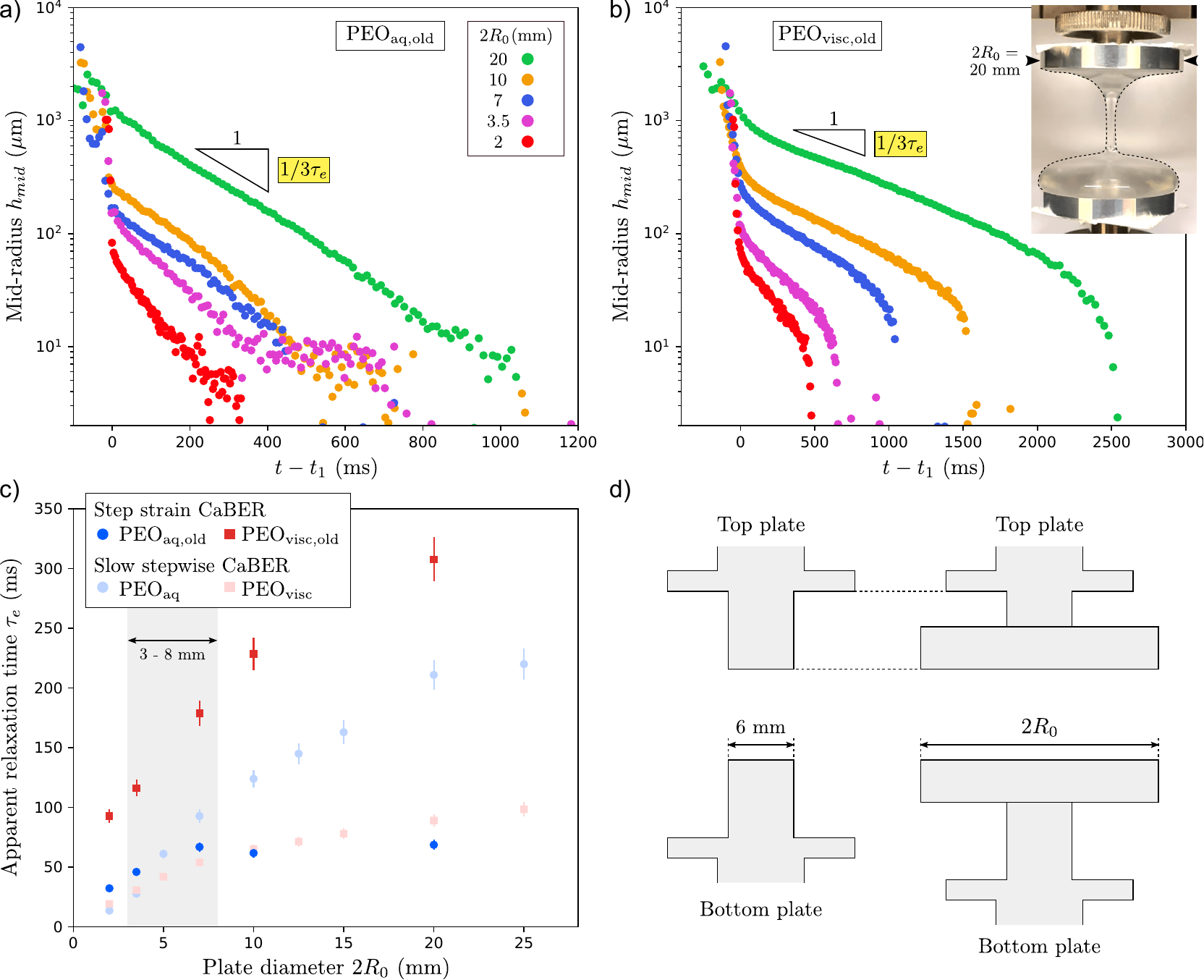}}
  \caption{(a,b) Time evolution of the mid-filament radius $h_{mid}(t)$ for the PEO$_{\mathrm{aq,old}}$ (a) and PEO$_{\mathrm{visc,old}}$ (b) solutions tested with plate diameters $2R_0 = 2$, $3.5$, $7$, $10$ and $20$~mm on a Haake CaBER-1 extensional rheometer. The time $t_1$ corresponds to the onset of the exponential regime. The legend is the same for both figures and the inset picture in (b) shows an example of thinning filament for the  PEO$_{\mathrm{visc,old}}$ solution tested with the $20$~mm diameter plates, the air-liquid interface being highlighted with dashed lines. (c) Apparent relaxation time $\tau_e$ against the plate diameter with, for reference, the values from figure \ref{fig:taue}(b) for the PEO$_{\mathrm{aq}}$ and PEO$_{\mathrm{visc}}$ solutions tested with the slow stepwise plate separation protocol described in \S\ref{subsec:Slow stepwise plate separation CaBER protocol}. The shaded area corresponds to the range of plate diameters $3-8$~mm available for a Haake CaBER-1 extensional rheometer. (d) Sketch of the original (left) and modified (right) design allowing for a change in plate diameter.}
\label{fig:Haake}
\end{figure}

\noindent To further test the universality of the dependence of the apparent relaxation time on the system size, we also performed experiments with a Haake CaBER-1 commercial extensional rheometer (Thermo Haake GmbH, Karlsruhe, Germany) with plate diameters $2R_0$ between $2$ and $20$~mm. This was achieved by sticking aluminium plates of prescribed diameters to the $6$~mm diameter plates provided with the rheometer where, as shown in figure \ref{fig:Haake}(d), the top plate was shortened to ensure that the total length would remain unchanged, allowing us to use the readings of the software to control the plate separation distance $L_p$. We chose an exponential plate separation profile of the form $L_p(t) = L_0 \exp{(\dot{\epsilon_0}t)}$ with initial and final separation distances $L_0$ and $L_f$ and with extension rate $\dot{\epsilon}_0 = t_f^{-1} \ln(L_f/L_0)$ where $t_f$ is the duration of the separation profile. Values of these parameters are shown in table \ref{tab:haake} for each plate diameter where $\Lambda_0 = L_0/R_0$ and $\Lambda_f = L_f/R_0$ are the initial and final aspect ratios. The initial gap is filled by a nearly cylindrical liquid bridge, yielding a liquid volume $V \approx \pi R_0^2 L_0$ and a non-dimensional liquid volume $V^* = V / R_0^3 \approx \pi \Lambda_0$ given in table \ref{tab:haake}. We choose $\Lambda_0 = 1$ for the smallest plates, consistent with Miller et al. \cite{miller2009effect}, and lower values for the largest plates for which, due to gravity, it was no longer possible to fit the sample in a $\Lambda_0 = 1$ initial gap. We keep the final-to-initial distance $L_f/L_0$ close to $4$ and choose the smallest available strike time $t_f = 20$~ms to maximise the extension rate $\dot{\epsilon}_0 \approx 69$~s$^{-1}$ and hence the Weissenberg number $\Wei_0 = \tau \dot{\epsilon}_0$ which, since the apparent relaxation times about to be discussed are larger than $30$~ms, is larger than $2$ and therefore within the range considered by Miller et al. \cite{miller2009effect}. This ensures that polymer chains do not relax during the (hence rightfully named) initial step-strain or step-stretch. Experiments of Miller et al. \cite{miller2009effect} for a fixed plate diameter $2R_0 = 3$~mm and initial aspect ratio $\Lambda_0 = 1$ showed that the (apparent) relaxation time $\tau_e$ (inferred from the exponential thinning regime) doesn't depend on the step-strain parameters $\Lambda_f$ (varied between $3$ and $15$) and $\Wei_0$ (varied between $0.5$ and $12$) for polymer solutions.

\begin{table}
\setlength{\tabcolsep}{4pt}
  \begin{center}
  \begin{footnotesize} 
\def~{\hphantom{0}}
  \begin{tabular}{cccccc}
  
    $2R_0$ (mm) & $\Lambda_0$  & $\Lambda_f$ & $t_f$ (ms) & $\dot{\epsilon}_0$ (s$^{-1}$) & $V^*$ \\[8pt]      
    2          & 1.00          & 3.98       & 20         & 69.3                           & 3.1 \\
    3.5        & 1.00          & 3.95       & 20         & 68.5                           & 3.2 \\
    7          & 1.00          & 3.97       & 20         & 69.0                           & 3.1 \\
    10         & 0.80          & 3.13       & 20         & 68.1                           & 2.5 \\
    20         & 0.41          & 1.64       & 20         & 69.3                           & 1.3 \\

  \end{tabular}
  \caption{Parameters of experiments with the Haake CaBER-1 commercial extensional rheometer.} 
  \label{tab:haake}
  \end{footnotesize} 
  \end{center}
\end{table}

We used the PEO$_{\mathrm{aq}}$ and PEO$_{\mathrm{visc}}$ solutions of figure \ref{fig:taue}(b) that we label here PEO$_{\mathrm{aq,old}}$ and PEO$_{\mathrm{visc,old}}$ since they were about 7 months old by the time we tested them on the Haake CaBER-1 rheometer compared to when they were tested with the slow stepwise plate separation protocol. The time evolution of the mid-filament radius $h_{mid}$ measured by the laser micrometer is shown in figures \ref{fig:Haake}(a) for the PEO$_{\mathrm{aq,old}}$ solution and in figure \ref{fig:Haake}(b) for the PEO$_{\mathrm{visc,old}}$ solution for all plate diameters. The relaxation time $\tau_e$ inferred from the exponential part of the thinning dynamics, calculated from the filament thinning rate $\vert \dot{h}/h \vert$ defined as $1/3 \tau_e$, is plotted against the plate diameter in figure \ref{fig:Haake}(c) for both liquids. We find that $\tau_e$ increases significantly as the plate diameter increases for the PEO$_{\mathrm{visc,old}}$ solution. For the PEO$_{\mathrm{aq,old}}$ solution, $\tau_e$ increases by a factor 2 for $2R_0$ between $2$ and $7$~mm and reaches a plateau for $2R_0 \ge 7$~mm. This initial increase is not caused by experimental error, as indicated by the small error bars estimated by repeating the experiment three times for each plate. These results confirm that the apparent relaxation time increases with the plate diameter regardless of whether the plates are separated slowly or rapidly in CaBER. The range of plate diameters $3-8$~mm used with the Haake CaBER-1 rheometer is shown in grey in figure \ref{fig:Haake}(c) to emphasise that, generally speaking, there is no reason to consider any apparent relaxation time measured in this range as `the' relaxation time. 

The apparent relaxation times of figure \ref{fig:taue}(b), measured with the slow stepwise plate separation CaBER protocol described in \S\ref{subsec:Slow stepwise plate separation CaBER protocol}, are shown in figure \ref{fig:Haake}(c) for reference. Values of $\tau_e$ are in general lower for the PEO$_{\mathrm{aq,old}}$ solution compared to the PEO$_{\mathrm{aq}}$ (fresh) one, consistent with the expected ageing of the solution over the 7 months separating the two sets of experiments which was confirmed by shear rheology experiments revealing a $29$\% decrease in $\eta_0$. On the other hand, values of $\tau_e$ are found to be higher for the PEO$_{\mathrm{visc,old}}$ compared to the PEO$_{\mathrm{visc}}$ (fresh) one. The (constant) shear viscosity was also found to be higher by about $24$\% which, since $\eta_s \gg \eta_p$, can be explained by an increase in the solvent viscosity which cannot be caused by evaporation since the solution was sealed. A possible explanation could be the development of microorganisms. The change in apparent relaxation times is therefore not necessarily caused by the change in plate separation protocol, consistent with Bazilevsky et al. \cite{bazilevsky1997failure} who found no significant difference in $\tau_e$ for solutions tested with both slow and fast protocols.

\section{Numerical method}
\label{appC}
The FENE-P model was solved with a variation of the method described by  Herrada \& Montanero \cite{herrada2016numerical}. The physical domains occupied by the liquid is mapped onto a rectangular domain through a coordinate transformation. Each variable and its spatial and temporal derivatives appearing in the transformed equations were written as a single symbolic vector. Then, we used a symbolic toolbox to calculate the analytical Jacobians of all the equations with respect to the symbolic vector. Using these analytical Jacobians, we generated functions that could be evaluated in the iterations at each point of the discretised numerical domains. 

The transformed spatial domain is discretised using $n_\eta=11$ Chebyshev spectral collocation points  in the transformed radial direction. We used $n_\xi=801$ equally spaced collocation points in the transformed axial direction $\xi$. The axial direction was discretised using fourth-order finite differences. Second-order backward finite differences were used to discretise the time domain. We used an automatic variable time step based on the norm of the difference between the solution calculated with a first-order approximation and that obtained from the second-order procedure. The nonlinear system of discretised equations was solved at each time step using the Newton method. The method is fully implicit.

\begin{small}
\bibliographystyle{elsarticle-num} 
\bibliography{bibliography.bib}
\end{small}

\end{document}